\documentclass[trackchanges,twocolumn,twocolappendix]{aastex701}

\usepackage{amsmath,threeparttable}
\usepackage{bm}

\begin{document}

\title{Interpreting ALMA Multiwavelength Continuum Observations of PDS~70~c:\\An Optically Thick Dust Ring in the Circumplanetary Disk}

\author[orcid=0000-0003-2993-5312,gname=Yuhito,sname='Shibaike']{Yuhito Shibaike}
\affiliation{Graduate School of Science and Engineering, Kagoshima University, 1-21-35 Korimoto, Kagoshima-shi, Kagoshima 890-0065, Japan}
\email[show]{yuhito.shibaike@sci.kagoshima-u.ac.jp}

\author[orcid=0000-0002-1886-0880,gname=Satoshi,sname=Okuzumi]{Satoshi Okuzumi}
\affiliation{Department of Earth and Planetary Sciences, Institute of Science Tokyo, 2-12-1 Ookayama, Meguro-ku, Tokyo 152-8551, Japan}
\email{okuzumi@eps.sci.titech.ac.jp}

\author[orcid=0000-0003-4902-222X,gname=Takahiro,sname='Ueda']{Takahiro Ueda}
\affiliation{Division of Science, National Astronomical Observatory of Japan, 2-21-1 Osawa, Mitaka-shi, Tokyo 181-8588, Japan}
\email{takahiro.ueda@nao.ac.jp}

\author[orcid=0000-0003-1958-6673,gname=Kiyoaki,sname=Doi]{Kiyoaki Doi}
\affiliation{Max-Planck-Institut f\"{u}r Astronomie, K\"{o}nigstuhl 17, 69117 Heidelberg, Germany}
\email{doi.kiyoaki.astro@gmail.com}

\author[orcid=0000-0003-1117-9213,gname=Misato,sname=Fukagawa]{Misato Fukagawa}
\affiliation{Astronomical Institute, Graduate School of Science, Tohoku University, 6-3 Aoba, Aramaki-aza, Aoba-ku, Sendai 980-8578, Japan}
\email{misato.fukagawa@astr.tohoku.ac.jp}


\begin{abstract}
Giant planets form small gas disks, called circumplanetary disks (CPDs), during gas accretion. The CPD of PDS 70 c has been detected by the Atacama Large Millimeter/submillimeter Array (ALMA) in (sub)millimeter continuum emission, which is interpreted as thermal emission from dust in the CPD. The resulting spectral index suggests that the disk is optically thick over a wide range of wavelengths. However, this is inconsistent with previous CPD dust models, which predict that the disk is optically thin because of radial dust drift. Here, we present a new interpretation of the multiwavelength observations: the CPD hosts an optically thick dust ring, whose existence has been discussed in the context of satellite formation. We demonstrate that a dust-ring model that incorporates gas accretion, dust evolution, and dust thermal emission, is consistent with the observations under reasonable conditions, whereas a conventional ring-less model requires more stringent conditions. We also show that the dust ring inferred from the observations potentially satisfies the conditions for exomoon formation via streaming instability and subsequent gravitational instability.
\end{abstract}

\keywords{\uat{Millimeter astronomy}{1061} --- \uat{Planet formation}{1241} --- \uat{Protoplanetary disks}{1300} --- \uat{Dust continuum emission}{412} --- \uat{Exoplanet formation}{492} --- \uat{Natural satellite formation}{1425}}


\section{Introduction} \label{sec:introduction}
Gas giant planets form small gas disks around them, called circumplanetary disks (CPDs), when they accrete gas from their protoplanetary disks (PPDs). CPD properties therefore provide key insights into the accretion process \citep[e.g.,][]{zhu15}.

Observations with the Atacama Large Millimeter/submillimeter Array (ALMA) have detected (sub)millimeter continuum emission from the gas-accreting planet PDS 70 c \citep{ise19,ben21}. This planet is one of two gas-accreting planets orbiting the young T Tauri star PDS~70, and both planets have also been detected in the infrared and in H$\alpha$ emission \citep[e.g.,][]{kep18,haf19}. The continuum emission from PDS~70~c has been observed in multiple ALMA bands. The emission was first detected in Band~7 ($855~\mu{\rm m}$) \citep{ise19,ben21} and subsequently in Bands~4 ($2.1~{\rm mm}$), 6 ($1.1~{\rm mm}$ and $1.4~{\rm mm}$), and 7 ($856~\mu{\rm m}$ and $873~\mu{\rm m}$) \citep{fas25,dom25}. In contrast, Band~3 ($3.0~{\rm mm}$) resulted in a nondetection in the original analysis \citep{doi24}, although a reanalysis yielded a marginal $2.6\sigma$ detection \citep{dom25}. A nondetection has been reported in Band~9 ($447~\mu{\rm m}$) \citep{dom25}. The radio emission from planet c has been interpreted as thermal emission from dust in the planet's CPD.

\begin{figure*}[ht!]
\centering
\includegraphics[width=0.65\linewidth]{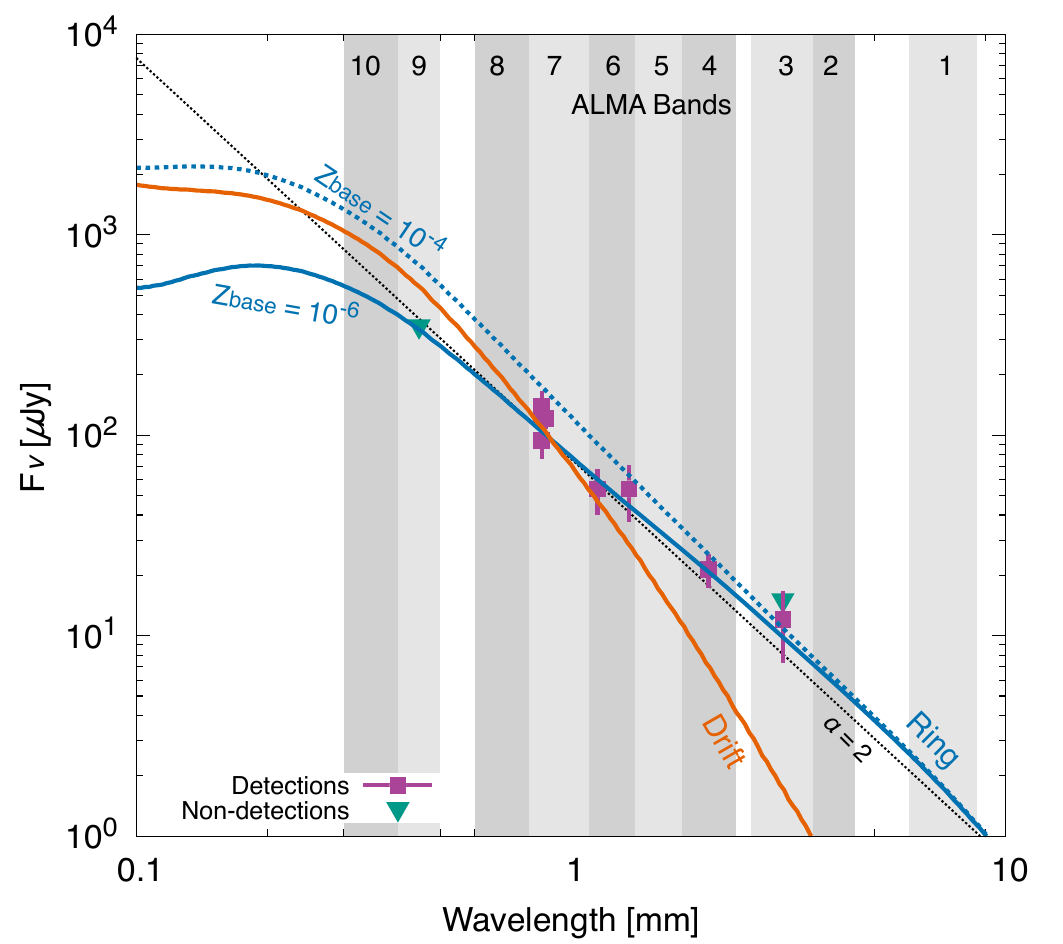}
\caption{Flux density of the continuum emission from PDS~70~c obtained from multiwavelength ALMA observations, together with model predictions for the dust thermal emission from the CPD in the Drift and Ring models. The squares and inverted triangles indicate detections and nondetections listed in Table \ref{tab:observations}. For the Ring model, the solid and dotted curves indicate $Z_{\rm base}=10^{-6}$ and $10^{-4}$, respectively.  The dotted line represents $\alpha=2$. The shaded bands represent the wavelength ranges of ALMA bands.
\label{fig:multiwavelength}}
\end{figure*}

The spatial resolution of the ALMA observations is insufficient to resolve the CPD, but the disk-integrated flux density has nevertheless provided information about the dust properties. \citet{shi24} proposed a model for dust evolution and emission in CPDs and constrained the planet mass and gas accretion rate of PDS~70~c based on the single wavelength observations at $855~\mu{\rm m}$ \citep{ben21}. However, the subsequent multiwavelength observations suggest that the disk shows an indication of optically thick dust emission in the spectral energy distribution (SED) slope \citep{fas25,dom25} (Figure \ref{fig:multiwavelength}; see the following sections for detailed descriptions). This is inconsistent with model predictions in which dust drifts inward in the CPD, leaving the disk optically thin, which has also been recognized as a general characteristic of PPDs \citep[e.g.,][]{tak02}. \cite{dom25} interpreted the optically thick spectrum and the Band~9 nondetection as H$_{\rm I}$ free-free continuum emission from an accretion shock at the CPD surface, rather than as dust thermal emission. \citet{cas22} and \citet{cas26} reported possible Band~7 flux variability and interpreted it in terms of the free–free emission as well.

In this work, we propose a novel interpretation in which the CPD of PDS 70 c is dominated by a ring-shaped, optically thick dust structure. This scenario is inspired by substructured PPDs whose dust emission is dominated by optically thick regions \citep[e.g.,][]{ric12,tri17}. Furthermore, a dust ring is expected to form in a CPD in the context of satellite formation \citep[e.g.,][]{dra18b} (see Section \ref{sec:ring-disk}). We investigate whether an optically thick dust ring in the CPD can explain the observations while comparing this scenario with a smooth dust disk undergoing radial drift.

\section{Methods} \label{sec:methods}
\subsection{Observational data} \label{sec:data}
We use data from ALMA observations analyzed in previous studies, as listed in Table \ref{tab:observations}. We refer to \citet{doi24}, \citet{fas25}, and \citet{dom25} for detailed descriptions. Here, we assume that the beam sizes are larger than the CPD size and hence the peak intensities of the detected continuum emission around PDS~70~c represent the CPD's integrated flux densities $F_{\nu}$, where $\nu$ is the frequency. In this work, we use the $3\sigma$ values as upper limits of the CPD emission in Bands~3 and 9 for the comparisons with the models. Figure \ref{fig:multiwavelength} shows the observed SED of the flux densities. Most of the detected flux densities follow a spectral slope $\alpha\equiv d\ln{F_{\nu}}/d\ln{\nu}$ close to 2. Note that we do not include the 2017 ALMA Band~7 dataset in our model comparisons, because the reported flux density for this epoch is sensitive to the adopted analysis \citep{cas22,cas26}. In Section \ref{sec:broad}, we use the data of Band~7 at $873~\mu{\rm m}$ observed on 2023-5-23, which was observed on the date closest to the Band~4 observation, 2024-4-30 \citep{dom25}. We obtain the disk-integrated spectral index between Bands~4 and 7, $\alpha_{\rm B4,B7}=2.01\pm0.19$.

\begin{table*}[htbp]
\caption{Multiwavelength ALMA observations of PDS~70~c}
\label{tab:observations}
\centering
\begin{tabular}{lllll}
\hline
Band & Wavelength [mm] & First date(s) of observations & Peak intensity $[\mu{\rm Jy}]$ & References \\
\hline\hline
3 & 3.07 & 2023-7-2 & $<14.82$ (nondetection; $3\sigma$) & \citet{doi24} \\
3 & 3.07& 2023-7-2 & $12.0\pm4.7$ (reanalysis) & \citet{dom25} \\
4 & 2.07 & 2024-4-30 & $21.4\pm4.1$ & \citet{dom25} \\
6 & 1.36 & 2020-3-6, 2021-7-14 & $54\pm17$ & \citet{fas25} \\
6 & 1.15 & 2020-3-2, 2021-5-26 & $54\pm14$ & \citet{fas25} \\
7 & 0.873 & 2023-5-23 & $121\pm13$ & \citet{dom25} \\
7 & 0.856 & 2016-8-14, 2019-7-27 & $94\pm18$ & \citet{fas25} \\
7 & 0.856 & 2021-7-18, 2021-8-6 & $139\pm28$ & \citet{fas25} \\
7 & 0.856 & 2023-3-2, 2023-6-1 & $127\pm23$ & \citet{fas25} \\
9 & 0.447 & 2023-6-11 & $<345$ (nondetection; $3\sigma$) & \citet{dom25} \\
\hline
\end{tabular}
\end{table*}

\subsection{Circumplanetary gas disk} \label{sec:gas}
Throughout this work, we model the CPD of PDS 70 c as a steady gas disk described by the ``gas-starved'' model, originally proposed by \citet{can02}. We adopt the updated version of this model from \citet{shi24} and \citet{shi25}, and summarize its basic framework below.

We calculate the radial distribution of the gas surface density $\Sigma_{\rm g}$ and midplane temperature $T$ of the CPD as functions of the planet mass $M_{\rm p}$ and the planet's gas accretion rate $\dot{M}_{\rm g}$. Gas is supplied from high altitudes in the parental PPD onto the CPD. Accretion through the CPD is driven by turbulence, although the turbulence in CPDs is expected to be weak \citep{fuj14}, which is parameterized by the dimensionless viscosity $\alpha_{\rm tur}$. The inner edge of the disk is truncated at $r_{\rm in}$ by the magnetospheric accretion, and the outer edge is truncated at $r_{\rm out}=R_{\rm H}/3$, where $R_{\rm H}$ is the Hill radius of the planet. The gas inflow region is $r_{\rm in}\leq r\leq r_{\rm c}$, where $r$ is the distance from the planet, and $r_{\rm c}$ is the centrifugal radius. The total gas mass flux of the inflow is then $\dot{M}_{\rm g,inf}=\dot{M}_{\rm g}/(1-4/5\sqrt{r_{\rm c}/r_{\rm out}})$. We estimate $r_{\rm c}$ as a function of the planet's Bondi radius and Hill radius based on previous hydrodynamical simulations of gas accretion \citep{war10}. The midplane gas temperature $T$ is determined by the balance between the viscous heating, irradiation from the PPD, irradiation from the planet, shock heating by the gas inflow, and radiative cooling from the disk surface. Irradiation from the PPD is represented as $T_{\rm PPD}$, which is treated as the minimum temperature of the CPD. This temperature determines the temperature of the outer part of the CPD, dominating the dust thermal emission. We describe the temperature structure of the disk in detail in Appendix \ref{sec:temperature}. See \citet{shi24} and \citet{shi25} for detailed descriptions of the disk model.

Note that our CPD model does not explicitly assume any formation scenario, although it remains unclear whether PDS 70 c formed through core accretion or gravitational instability \citep[e.g.,][]{cri23}. Our model is based on general physical ingredients of disk structure and accretion rather than on a specific planet-formation pathway. Hydrodynamical simulations also show that CPDs can form in both scenarios \citep{szu17a}.

\subsection{Dust distribution in the CPD: Drift or Ring} \label{sec:dust}
We consider two models for the dust distribution in the CPD: the Drift and Ring models. In both models, we calculate the steady radial distributions of the dust surface density, $\Sigma_{\rm d}$, and the maximum particle radius, $a_{\rm max}$. We take the internal density $\rho_{\rm int}$ to be $1.4~{\rm g~cm}^{-3}$ for icy particles (outside the snowline) and $3.0~{\rm g~cm}^{-3}$ for rocky particles (inside the snowline).

The Drift model, as adopted by \citet{shi24} and \citet{shi25}, computes $\Sigma_{\rm d}$ and $a_{\rm max}$ by accounting for the particles' coagulation, fragmentation, and radial drift toward the planet. Small dust particles are supplied with the gas inflow to an annulus of $r_{\rm in}\leq r\leq r_{\rm c}$. The dust-to-gas inflow mass flux ratio $x\equiv\dot{M}_{\rm d,inf}/\dot{M}_{\rm g,inf}$ is a key parameter controlling dust evolution, where $\dot{M}_{\rm d,inf}$ is the total dust mass flux of the inflow.

The Ring model assumes that dust particles are concentrated at a specific distance from the planet. The dust surface density in this model is given by
\begin{equation}
\Sigma_{\rm d}=\Sigma_{\rm base}+\Sigma_{\rm peak}\exp\left\{-\dfrac{(r-r_{\rm ring})^{2}}{2w_{\rm ring}^{2}}\right\},
\label{l}
\end{equation}
where $\Sigma_{\rm base}$ and $\Sigma_{\rm peak}$ are the base and peak dust surface densities, and  $r_{\rm ring}$ and $w_{\rm ring}$ are the peak position and width of the ring, respectively. We take the base and peak dust-to-gas ratios $Z_{\rm base}=\Sigma_{\rm base}/\Sigma_{\rm g}$ and $Z_{\rm peak}=\Sigma_{\rm peak}/\Sigma_{\rm g}$, as free parameters of the model. In the Ring model, we consider that dust particles grow large by coagulation until fragmentation limits their growth. The maximum particle radius is then
\begin{equation}
a_{\rm max}=\dfrac{\Sigma_{\rm g}}{1.15\pi\alpha_{\rm tur}\rho_{\rm int}}\left(\dfrac{v_{\rm frag}}{c_{\rm s}}\right)^{2},
\label{amax}
\end{equation}
where $v_{\rm frag}$ is the critical fragmentation speed of dust \citep{orm07,bir09,oku19}. The isothermal sound speed of the CPD is $c_{\rm s}=\sqrt{k_{\rm B}T/m_{\rm g}}$, where $k_{\rm B}$ and $m_{\rm g}$ are the Boltzmann constant and the mean molecular mass of the gas, respectively.

\subsection{Thermal emission from dust} \label{sec:emission}
The dust thermal emission is calculated from $\Sigma_{\rm d}$ and $a_{\rm max}$ using the opacity module OpTool \citep{dom21}. We adopt the DSHARP dust opacity for the optical properties of the particles \citep{bir18}. We assume a power-law dust size distribution with a slope of $-3.5$ and a fixed minimum dust radius of $0.1~\mu{\rm m}$.

The emergent intensity $I_\nu$ is calculated using Equation (3) of \citet{sie24}, which assumes a vertically isothermal disk and accounts for both absorption and scattering. We assume that the dust temperature is equal to the midplane gas temperature $T$. We define the local spectral index by $\alpha_{\rm loc}\equiv d\ln{I_{\nu}}/d\ln{\nu}$. The disk-integrated flux density from the dust in the CPD is
\begin{equation}
F_{\nu}=\dfrac{2\pi\cos{i}}{d^{2}}\int^{r_{\rm out}}_{r_{\rm in}}I_{\nu}rdr.
\label{Fnu}
\end{equation}

\subsection{Parameters} \label{sec:parameters}
For both the Drift and Ring models, we first calculate a fiducial case to investigate the basic properties of the models with representative parameter values listed in the third column of Table \ref{tab:parameters}. We fix $M_{\rm p}$, $\dot{M}_{\rm g}$, $T_{\rm PPD}$, $\alpha_{\rm tur}$, and $v_{\rm frag}$ based on previous studies (\citet{shi24,sie25,fuj14,liu26}; see Appendix \ref{sec:validity} for detailed descriptions). Values of the other parameters are uncertain. In the Drift model, we choose $x=4.7\times10^{-3}$ to reproduce the observed flux density in Band~7 ($100~\mu{\rm Jy}$). In the Ring model, we assume $Z_{\rm peak}=1$ and $r_{\rm ring}=r_{\rm c}$ for simplicity. We consider two values of $Z_{\rm base}$, $10^{-6}$ and $10^{-4}$, to assess the contribution from the regions outside the ring. We then assume $w_{\rm ring}=97~R_{\rm J}$ that reproduces the observed flux density in Band~7. We discuss the validity of the values of $x$ and $Z_{\rm peak}$ in Sections \ref{sec:thick-disk} and \ref{sec:ring-disk}.

We then explore a broader parameter space to assess the potential impact of the parameters. We perform $n=2000$ calculations for each of the Drift and Ring models, with the parameters randomly varied within the ranges listed in the fourth column of Table \ref{tab:parameters} (see Sections \ref{sec:thick-disk} and \ref{sec:ring-disk}, and Appendix \ref{sec:validity} for their validity).

The other properties of the planet and its host system are fixed throughout this work. The host-star mass is $M_{*}=0.76~M_{\odot}$ \citep{mul18}, and the distance to the system is $d=112.4~{\rm pc}$ \citep{gaia23}. The orbital distance of PDS~70~c is $a_{\rm pl}=34.5~{\rm au}$ \citep{haf19}. The radius and effective temperature of planet c are estimated from infrared observations as $R_{\rm pl}=2.0~R_{\rm J}$ and $T_{\rm pl,eff}=1051~{\rm K}$, respectively \citep{wan21}. We also assume the inclination of the CPD is the same as that of the PPD, $i=51.7^{\circ}$ \citep{kep19}.

\begin{table*}[htbp]
\caption{Parameter values}
\label{tab:parameters}
\centering
\begin{tabular}{llll}
\hline
Description & Symbol & Fiducial case & Broad-parameter cases \\
\hline\hline
Planet mass & $M_{\rm p}$ & $10~M_{\rm J}$ & $4$--$12~M_{\rm J}$ \\
Planet's gas accretion rate & $\dot{M}_{\rm g}$ & $2\times10^{-7}~M_{\rm J}~{\rm yr^{-1}}$ & ($0.4$--$22$)$\times10^{-8}~M_{\rm J}~{\rm yr^{-1}}$ \\
PPD Temperature & $T_{\rm PPD}$ & $22~{\rm K}$ & $10$--$50~{\rm K}$ \\
Turbulence strength in CPD & $\alpha_{\rm tur}$ & $10^{-4}$ & $10^{-6}$--$10^{-2}$ \\
Critical fragmentation speed & $v_{\rm frag}$ & $1~{\rm m~s^{-1}}$ & $0.3$--$50~{\rm m~s^{-1}}$ \\
\hline
Drift model \\
\hline
Inflow dust-to-gas mass flux & $x$ & $4.7\times10^{-3}$ & $10^{-4}$--$1$ \\
\hline
Ring model \\
\hline
Peak Dust-to-gas surface density ratio & $Z_{\rm peak}$ & $1$ & $10^{-4}$--$1$ \\
Base dust-to-gas surface density ratio & $Z_{\rm base}$ & $10^{-6},10^{-4}$ & $10^{-8}$--(given value of $Z_{\rm peak}$) \\
Ring location & $r_{\rm ring}$ & $r_{\rm c}$ & $0.1$--$1~r_{\rm out}$ \\
Ring width & $w_{\rm ring}$ & $97~R_{\rm J}~(0.037~r_{\rm ring})$ & $0.01$--$0.3~r_{\rm ring}$ \\
\hline
\end{tabular}
\end{table*}

\section{Results} \label{sec:results}
\subsection{Fiducial case} \label{sec:fiducial}
\begin{figure*}[ht!]
\centering
\includegraphics[width=0.95\linewidth]{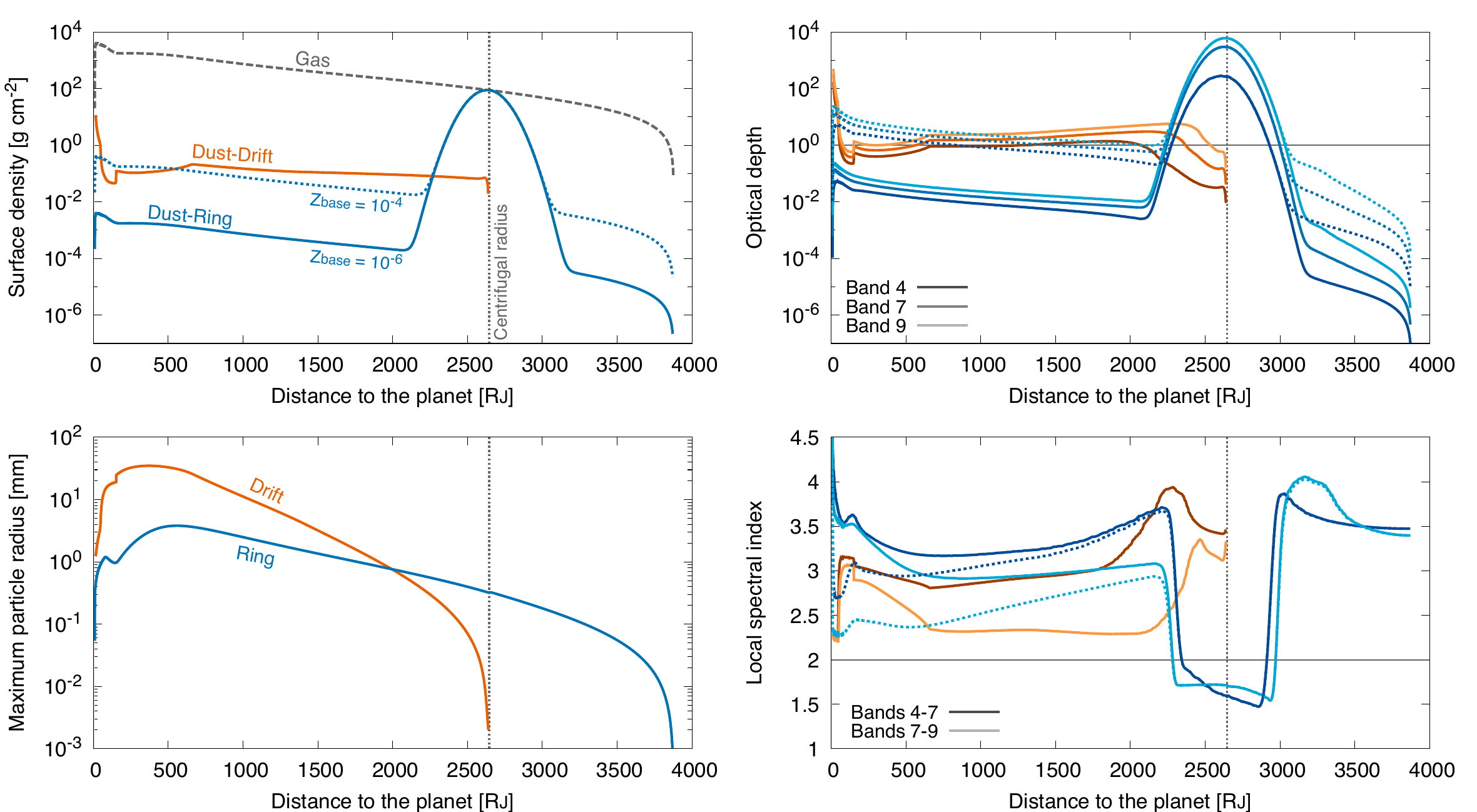}
\caption{Radial distribution of the dust properties in the CPD of PDS~70~c in the Drift and Ring models for the fiducial case. For the Ring model, the solid and dotted curves indicate $Z_{\rm base}=10^{-6}$ and $10^{-4}$, respectively. The different shades of each color represent the ALMA bands in the right two panels. The dashed curve in the left upper panel is the gas surface density of the CPD, used in both Drift and Ring models. The vertical dotted lines indicate the centrifugal radius.
\label{fig:models}}
\end{figure*}

We start with the fiducial case, where the parameters are listed in the third column of Table \ref{tab:parameters}. Figure \ref{fig:models} shows the basic properties of the dust in the CPD.

The left panels of Figure~\ref{fig:models} show the surface density and maximum particle radius profiles. In the Drift model, dust exists only inside the centrifugal radius of $r_{\rm c}=2645~R_{\rm J}$. This is because the dust particles are supplied inside $r_{\rm c}$ and drift inward as they grow through mutual collisions. The maximum particle radius is determined by radial drift in most regions, with fragmentation playing a role only at $r\lesssim1000~R_{\rm J}$. In the Ring model, the base dust surface density in the $Z_{\rm base}=10^{-4}$ case is close to $\Sigma_{\rm d}$ in the Drift model. The maximum particle radius in the Ring model, which is set by fragmentation, is about $0.3~{\rm mm}$ in the ring.

The right panels of Figure~\ref{fig:models} show the optical depth and spectral index distributions. In the Drift model, the outer edge of the dust disk (i.e., $r\approx r_{\rm c}$) yields $\tau_{\nu}<1$ at all bands, resulting in a local spectral index of $\alpha_{\rm loc}\approx3$--$4$. At $r\lesssim2000~R_{\rm J}$, the local spectral index is smaller, with $\alpha_{\rm loc}\approx2.3$ for Bands~7 and 9 and $\alpha_{\rm loc}\approx3$ for Bands~4 and 7. This is because the particles continue growing as they drift inward (see the left lower panel). In the Ring model with $Z_{\rm base}=10^{-6}$, in contrast, the optical depth exceeds unity and $\alpha_{\rm loc}\approx2$ only within the ring. Note that $\alpha_{\rm loc}$ is even smaller than two owing to the deviation from the Rayleigh--Jeans law. In the Ring model with $Z_{\rm base}=10^{-4}$, the local spectral indices outside the ring are similar to those in the $Z_{\rm base}=10^{-6}$ case, but are smaller inside the ring owing to the higher optical depth.

Figure \ref{fig:multiwavelength} compares the SED constructed from the observed peak intensities of PDS~70~c with the model predictions of the disk-integrated flux densities. The figure shows that the spectral slope of the Drift model is too steep to be consistent with the observations at both longer and shorter wavelengths than Band~7, where $\alpha\approx3$--$4$. This is because $\alpha_{\rm loc}\approx3$--$4$ at the outer part of the dust disk (Figure \ref{fig:models}). Because the surface area of a fixed-width annulus increases with radius, the disk-integrated flux density of the dust thermal emission is dominated by the emission from the outer part.

In contrast, the Ring model with $Z_{\rm base}=10^{-6}$ has a spectral slope compatible with the observations at all wavelengths. From Bands~7 to 3 (and the longer wavelengths), $\alpha$ remains close to two. This is because $\alpha_{\rm loc}\approx2$ in the ring, where its emission dominates the disk-integrated flux. This model is also consistent with the nondetection in Band~9. Because the temperature is sufficiently low that the Rayleigh--Jeans approximation does not hold, $\alpha$ is smaller than two at short wavelengths. Here, the temperature at the ring is nearly equal to the PPD temperature of $T_{\rm PPD}=22~{\rm K}$ (see Appendix \ref{sec:temperature}).

However, in the case of $Z_{\rm base}=10^{-4}$, $\alpha$ is larger than two at short wavelengths. This is because the emission from the outside of the dust ring, where $\alpha_{\rm loc}\approx3$--$4$, also contributes to the disk-integrated flux density (see Figure \ref{fig:models}). This result indicates that the CPD likely has a highly concentrated dust ring.

\subsection{Broad-parameter cases} \label{sec:broad}
\begin{figure*}[ht!]
\centering
\includegraphics[width=0.95\linewidth]{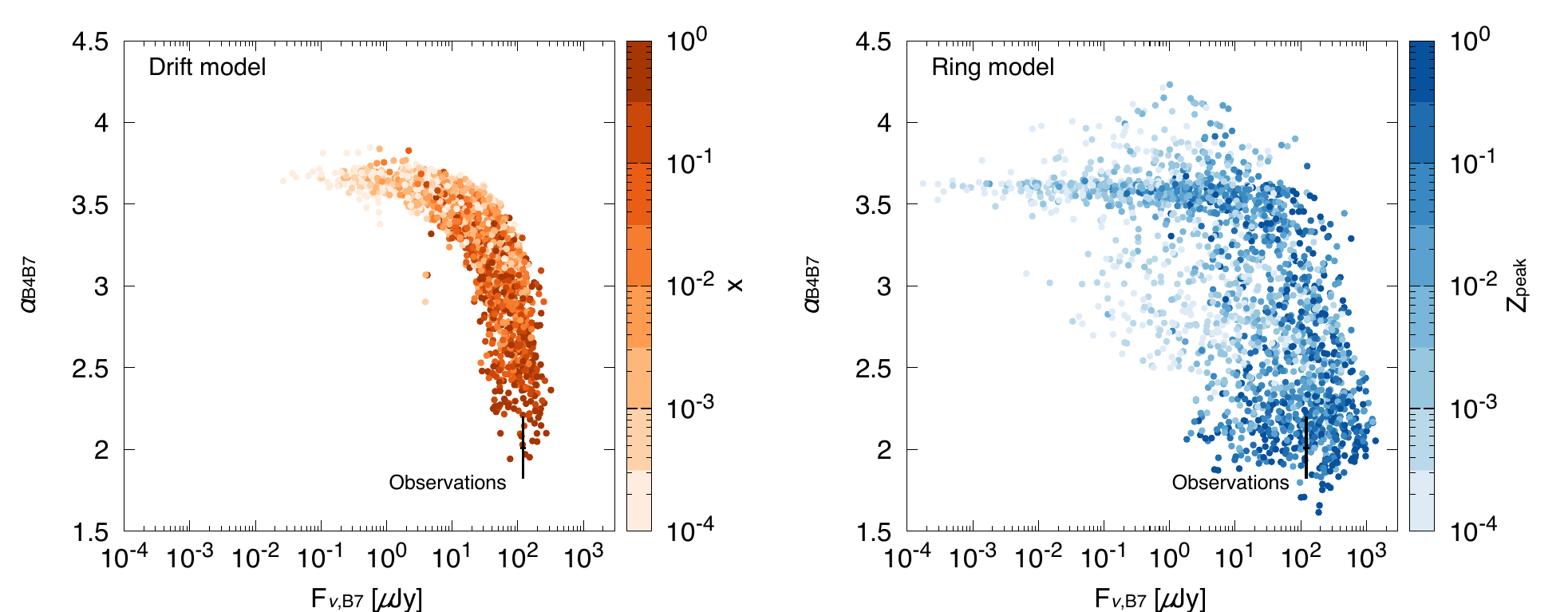}
\caption{Comparisons of the predictions by the Drift model (left panel) and Ring model (right panel) with ALMA observations of PDS~70~c. The parameter ranges of the models are shown in the ``Broad-parameter cases'' column in Table \ref{tab:parameters}. The large cross represents the disk-integrated flux density in ALMA Band~7 ($F_{\nu,{\rm B7}}=121\pm13~\mu{\rm Jy}$) and the disk-integrated spectral index of Bands~4 and 7 ($\alpha_{\rm B4,B7}=2.01\pm0.19$) observed by \citet{dom25}. The colors in the left and right panels represent the dust-to-gas mass ratio in the gas flow, $x$, and the dust-to-gas surface density ratio at the peak, $Z_{\rm peak}$, respectively.
\label{fig:comparisons}}
\end{figure*}

We now investigate cases in which the parameters are varied over the broad ranges listed in the fourth column of Table \ref{tab:parameters}. Figure \ref{fig:comparisons} compares predictions for $F_{\nu,{\rm B7}}$ and $\alpha_{\rm B4,B7}$ from the Drift and Ring models with the observations by \citet{dom25}.

In the Drift model (left panel of Figure \ref{fig:comparisons}), only calculations with high dust-to-gas mass ratios in the inflow, $x>0.1$, are consistent with $\alpha_{\rm B4,B7}=2.01\pm0.19$. In contrast, the Ring model reproduces the observed spectral index over a wide range of $Z_{\rm peak}$ above $10^{-2.5}\approx3\times10^{-3}$, with larger $Z_{\rm peak}$ values tending to be more consistent with the observed $\alpha_{\rm B4,B7}$ (right panel of Figure \ref{fig:comparisons}).

\section{Discussion} \label{sec:discussion}
\subsection{Feasibility of optically thick CPD formation without a ring} \label{sec:thick-disk}
In Section \ref{sec:broad}, we showed that the Drift CPD model requires high dust-to-gas inflow ratios of $x>0.1$ to explain the observed SED feature indicative of optically thick emission. However, this condition is not easily achieved. The ALMA images show that the region surrounding PDS~70~c is depleted in dust that emits at (sub-)millimeter wavelengths \citep[e.g.,][]{ben21}. In contrast, the images taken by the James Webb Space Telescope (JWST) suggest that there could be a ``bridge'' (i.e., inflow) containing some amount of smaller dust involved in infrared emission, which connects the CPD and the outer bright ring of the PPD \citep{chr24}. However, previous studies have argued that the amount of small dust in such inflow is small, since the gas is supplied from high altitudes in the parental PPD, while the dust particles settle to the midplane. For example, \citet{mae24} showed by local hydrodynamical and dust simulations that $x\lesssim10^{-2}$ even under the most favorable conditions considered in that study. Global hydrodynamical and dust simulations by \citet{szu22} showed that planets can stir dust up from the midplane to high altitudes, and the dust-to-gas mass ratio supplied to the Hill sphere of the planet can reach $\sim0.1$ in some situations. However, their numerical resolution is insufficient to resolve dust motions inside the Hill sphere, making it difficult to derive an exact value of $x$ from their simulations. Dust could also be supplied by the ablation of planetesimals captured by the CPD \citep{ron20}, but the amount of dust produced depends on how many planetesimals are transported to the vicinity of the planet's orbit, which is also highly uncertain.

\subsection{Feasibility of dust-ring formation in a CPD} \label{sec:ring-disk}
Compared to optically thick smooth CPDs in the Drift models, disks containing optically thick rings appear more plausible, as such structures have been predicted in previous theoretical studies in the context of satellite formation. Many hydrodynamical simulations have shown that a radial outflow can occur at the midplane of a CPD \citep[e.g.,][]{tan12,bat18}, which can trap dust particles at a specific distance from the central planet \citep{dra18b}. The peak dust-to-gas surface density ratios (i.e., $Z_{\rm peak}$) are higher than $0.1$ in all their calculations. The location of the dust ring is determined by the balance between the inward radial drift and the outward advection by gas.

Another possible mechanism for forming a dust ring is dust trapping at a gas-pressure bump, which has been extensively studied in the context of PPDs \citep[e.g.,][]{dul18}. Gas and dust hydrodynamical simulations showed that such a bump can trap dust until $Z_{\rm peak}$ reaches unity \citep{kan18b} when there is a continuous dust supply to the bump. \citet{shi23b} showed that magnetic wind-driven accretion can occur in the inner part of CPDs. The gas surface density is then reduced only at the inner part, resulting in the formation of a gas-pressure bump at the outer boundary of the wind-driven accretion region. Furthermore, if a large satellite with the pebble isolation mass is embedded in the CPD, a gas-pressure bump could also be created by that exomoon. It has indeed been shown that Ganymede could have reached the pebble isolation mass in the circum-Jovian disk \citep{shi19,ron20}.

If dust particles supplied to the CPD are efficiently accumulated in the ring, the equivalent value of $x$ required for the Ring model, ${\bm x_{\rm eq}}$, could be much smaller than ${\bm x}$ of the Drift model. The dust mass in the CPD including the dust ring is $3.1~M_{\rm E}$ in the fiducial case ($Z_{\rm peak}=1$). This dust mass would be supplied over $5.4~{\rm Myr}$ for ${\bm x_{\rm eq}}=3\times10^{-3}$, given that the total gas inflow rate $\dot{M}_{\rm g,inf}=1.9\times10^{-4}~M_{\rm E}~{\rm yr}^{-1}$ in the fiducial case (see also Appendix \ref{sec:equivalent}). Here, the age of the central star PDS~70 is estimated to be $5.4\pm1.0~{\rm Myr}$ \citep{mul18}, although the age of PDS~70~c and its accretion period are unknown.

\subsection{Satellite formation in the dust ring} \label{sec:satellite}
We now ask whether satellites can form in such an optically thick dust ring. The streaming instability (SI) typically requires $\rho_{\rm d}/\rho_{\rm g} \gtrsim 0.4$ for small particles with ${\rm St}\lesssim 10^{-2}$ \citep{li21}, where $\rho_{\rm d}$ and $\rho_{\rm g}$ are the dust and gas densities at the midplane, respectively, and ${\rm St}=\pi\rho_{\rm int}a_{\rm max}/(2\Sigma_{\rm g})$ is the Stokes number of the largest particles. Our fiducial Ring model predicts ${\rm St}\approx8\times10^{-4}$ at the ring peak. Since the dust-to-gas density ratio at the ring peak is given by $\rho_{\rm d}/\rho_{\rm g}|_{\rm peak}\approx Z_{\rm peak}\sqrt{1+{\rm St}/\alpha_{\rm tur}}$ \citep{you07}, the dust ring in the fiducial case satisfies the SI condition when $Z_{\rm peak}\gtrsim0.1$. We note that the SI criterion adopted here assumes smooth PPDs, and \citet{car21} showed that it is not necessarily stringent in a gas-pressure bump.

Further concentration of dust into satellitesimals via gravitational instability (GI) occurs if the dust density of a clump becomes larger than the Roche density \citep[e.g.,][]{sek98}. The gas density at the ring midplane is estimated to be $\rho_{\rm g}=\Sigma_{\rm g}/(\sqrt{2\pi}H_{\rm g})=1.6\times10^{-11}~{\rm g~cm}^{-3}$, where $H_{\rm g}=c_{\rm s}/\Omega_{\rm K}$ is the gas scale height of the CPD, and $\Omega_{\rm K}$ is the Keplerian frequency. In PPDs, the dust density inside clumps formed by SI can increase to $\rho_{\rm cl}\sim10^{3}\rho_{\rm g}$ \citep{bai10}. If this also holds in CPDs, the dust density inside SI-induced clumps in the ring would be $\rho_{\rm cl}\sim10^{-8}~{\rm g~cm}^{-3}$. On the other hand, the Roche density at the ring is $\rho_{\rm R}=9M_{\rm p}/(4\pi r^{3})=2.0\times10^{-9}~{\rm g~cm}^{-3}$. Therefore, the predicted dust density of the clumps is larger than the Roche density. This is a notable result because achieving the GI-onset condition, $\rho_{\rm cl}>\rho_{\rm R}$, is difficult in the circum-Jovian disk, owing to its lower gas density and smaller size compared with the CPD of PDS~70~c \citep{dra18b}. Considering that protoplanets can form from a planetesimal ring in a PPD \citep[e.g.,][]{han09}, it is reasonable to expect that satellites can form from such a satellitesimal ring in the CPD of PDS~70~c.

\subsection{The free-free emission interpretation} \label{sec:free-free}
Our CPD model neither supports nor rules out a possible contribution of the free-free emission to the continuum from PDS~70~c proposed by \citet{dom25}. Our model calculates only the dust thermal emission and does not model the free-free emission. One possible interpretation of the observations is a mixed origin of the two types of emission. In such a scenario, the free-free component could also contribute to the possible time variability reported in the Band~7 observations \citep{cas22,cas26}. However, the H$_{\rm I}$ free-free emission from the disk surface accretion shocks, identified by \citet{dom25} as the most viable component, is not expected to occur under the conditions considered in this study. In their two calculations, the adopted accretion rates onto the CPD are more than three orders of magnitude higher than those adopted in this study. Moreover, such high accretion rates are inconsistent with the estimates from the previous observations of PDS~70~c (see Appendix \ref{sec:validity}) unless one assumes an extreme situation in which only $\sim0.1\%$ of the gas accreted onto the disk is subsequently accreted to the planet.

It is challenging to distinguish the two origins using SED-based approaches, since the two emission mechanisms could produce similar SED slopes. However, future observations by the Next Generation Very Large Array (ngVLA) will be able to distinguish the two origins. The free-free emission only arises from the very inner part of the CPD \citep{dom25}, while the dust thermal emission is dominated by the outer part of the disk. \citet{shi25c} show that ngVLA Band~6 ($3~{\rm mm}$) can detect the continuum emission from PDS~70~c and spatially resolve the dust thermal emission of the possible substructures of the CPD such as a dust ring.

\section{Conclusions} \label{sec:conclusions}
In order to explain the recent multiwavelength (sub-)millimeter continuum ALMA observations \citep{fas25,dom25}, we have presented the hypothesis that the CPD of PDS~70~c hosts an optically thick dust ring. We found that a high dust-to-gas mass ratio of the inflow onto the CPD is required to be consistent with the observations in a conventional smooth dust disk, although, theoretically, the inflow should be dust-depleted \citep[e.g.,][]{mae24}. In contrast, our dust-ring model accounting for gas accretion and dust evolution is consistent with the observations over a reasonable range of parameter values. Therefore, we conclude that the presence of an optically thick dust ring in the CPD of PDS~70~c provides a plausible explanation for the observations. This interpretation supports the presence of either a midplane gas outflow or a gas-pressure bump within the CPD, offering a new observational clue to the gas-accretion process of giant planets. Furthermore, the dust ring potentially satisfies the conditions for SI and subsequent GI, suggesting formation of exomoons around PDS~70~c.

\begin{acknowledgments}
We thank the referee for constructive comments. We thank Akimasa Kataoka, Benedikt Gottstein, and Gabriel-Dominique Marleau for very useful discussions. This work was supported by JSPS KAKENHI grant Nos. JP22H01274 and JP24K22907.
\end{acknowledgments}





%
\facilities{ALMA}

\software{OpTool \citep{dom21}}


\appendix
\section{Validity of the parameter values} 
\label{sec:validity}
Here, we show the validity of the parameter values listed in Table \ref{tab:parameters}. In the fiducial case, we adopt the representative parameter values. We fix the planet mass and gas accretion rate at the fiducial values in \citet{shi24}. We fix the PPD temperature using Band~9 observations of the PPD. The optically thick outer ring at $R=67.4~{\rm au}$ has a peak brightness temperature of $15.5~{\rm K}$, which likely traces the physical temperature \citep{sie25}. Assuming $T_{\rm PPD}\propto R^{-1/2}$, we obtain $T_{\rm PPD}=22~{\rm K}$ at the orbit of PDS 70 c. We assume weak turbulence of dimensionless viscosity $\alpha_{\rm tur}=10^{-4}$, considering that the magnetorotational instability is unlikely to occur in CPDs \citep{fuj14}. We fix the critical fragmentation speed at the value recently inferred from ALMA polarization observations of dust in the ring of the parental PDS 70 disk \citep{liu26}.

In the broad-parameter cases, the ranges of the planet mass and the gas accretion rate are based on the planet's K$-$L color, H$\alpha$ line width, and evolution models \citep{haf19}. We vary the PPD temperature and the turbulence strength over relatively broad ranges, since the previous estimates include uncertainty \citep{zhu16,law24}. The range of the critical fragmentation velocity is based not only on observations but also on numerical simulations of collisions \citep{wad13,liu26}. To assess the impacts of these parameters, we vary $x$ and $Z_{\rm peak}$ over broad ranges of $10^{-4}$--$1$ (see Sections \ref{sec:thick-disk} and \ref{sec:ring-disk}). Here, the minimum and maximum values of $Z_{\rm peak}$ are consistent with the fiducial values of $Z_{\rm base}$ and $Z_{\rm peak}$, respectively. We then vary $Z_{\rm base}$ from $10^{-8}$ to the given $Z_{\rm peak}$ value. We also vary $r_{\rm ring}$ and $w_{\rm ring}$ over broad ranges, since there have been no constraints on the locations and widths of dust rings in CPDs.

\section{Temperature structure of the CPD}
\label{sec:temperature}
In this work, we determine the radial distribution of the midplane temperature of the CPD, $T(r)$, by calculating the balance between the heating and cooling (see \citet{shi24} and \citet{shi25} for detailed descriptions). The heat sources are the viscous heating, irradiation from the surrounding PPD (described by $T_{\rm PPD}$), irradiation from the planet, and shock heating on the CPD's surface by the gas inflow. For the irradiation from the planet, we consider the irradiation on the disk surface and the direct irradiation through the midplane. For the cooling, we only consider the radiative cooling from the disk surface.

\begin{figure}[ht!]
\centering
\includegraphics[width=0.95\linewidth]{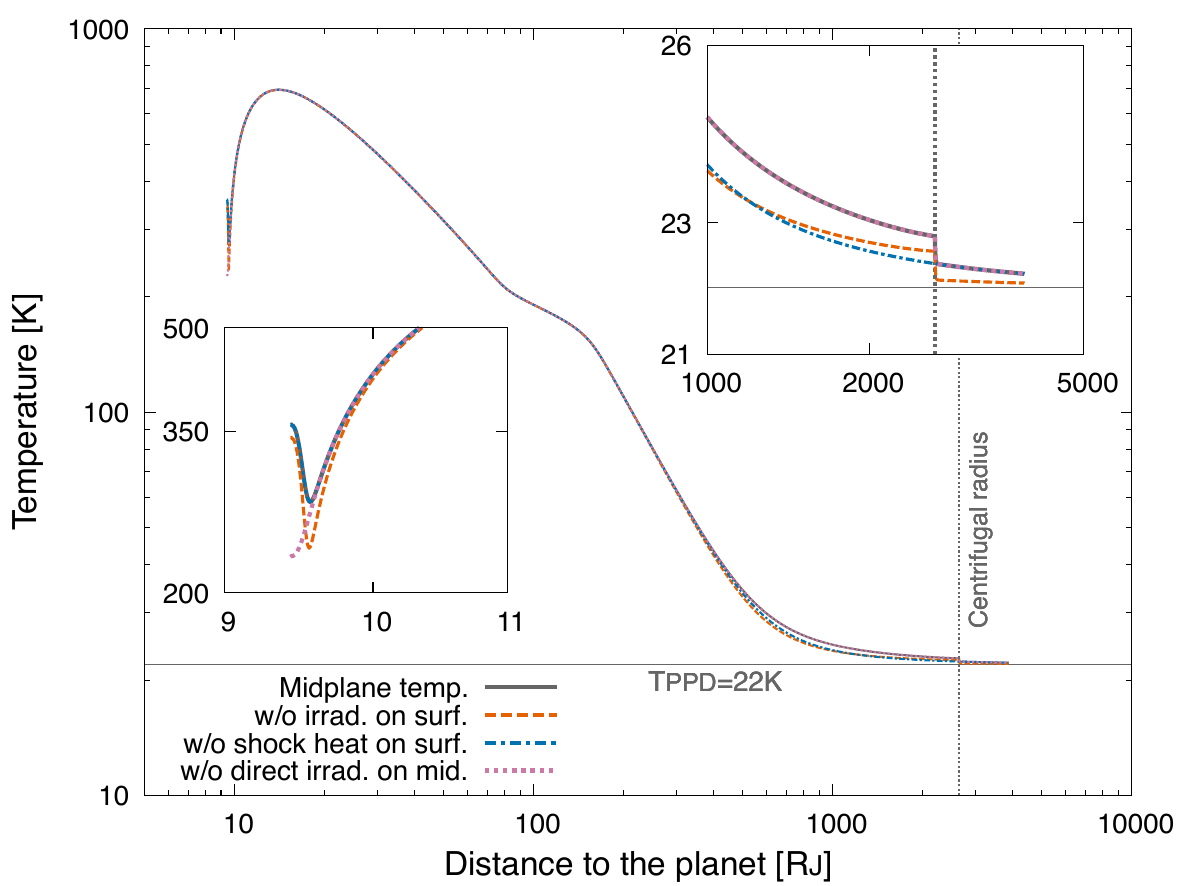}
\caption{Radial distribution of the midplane temperature and profiles calculated without the various heat sources. The small panels represent the enlarged views of the profiles at the inner and outer parts of the disk. The solid curves behind the others are the midplane temperature $T$. The dashed, dashed-dotted, and dotted curves represent the temperature without the irradiation from the planet on the CPD surface, the shock heating on the surface by the gas inflow, and the direct irradiation from the planet through the midplane. The horizontal line is the PPD temperature surrounding the CPD, $T_{\rm PPD}=22~{\rm K}$. The dotted vertical lines represent the centrifugal radius, $r_{\rm c}=2645~R_{\rm J}$.
\label{fig:temperature}}
\end{figure}

Figure \ref{fig:temperature} shows the radial distribution of the midplane temperature in the fiducial case (third column of Table \ref{tab:parameters}), together with profiles calculated without the various heat sources. The figure shows that the temperature around the centrifugal radius is dominated by the irradiation from the PPD ($T_{\rm PPD}=22~{\rm K}$), although the irradiation from the planet on the disk surface slightly increases $T$ at the outer part of the disk ($r\gtrsim300~R_{\rm J}$). The shock heating also only slightly increases the temperature at $r\leq r_{\rm c}$ (i.e., the gas inflow region). The direct irradiation from the planet through the midplane increases the temperature only at the very inner part of the CPD ($r\lesssim10~R_{\rm J}$). Except in the outer part of the disk, viscous heating is the dominant heat source in the CPD. However, the viscous heating has little effect on the disk-integrated flux density, since the disk-integrated flux is dominated by the flux from the outer part. Therefore, the disk-integrated flux density of the dust thermal emission from the CPD is determined by the irradiation from the PPD, that is, the temperature of the PPD surrounding the CPD, under the assumption that the dust and gas have the same temperature.

\section{Equivalent dust-to-gas ratios in the Drift and Ring models}
\label{sec:equivalent}
\begin{figure*}[ht!]
\centering
\includegraphics[width=0.95\linewidth]{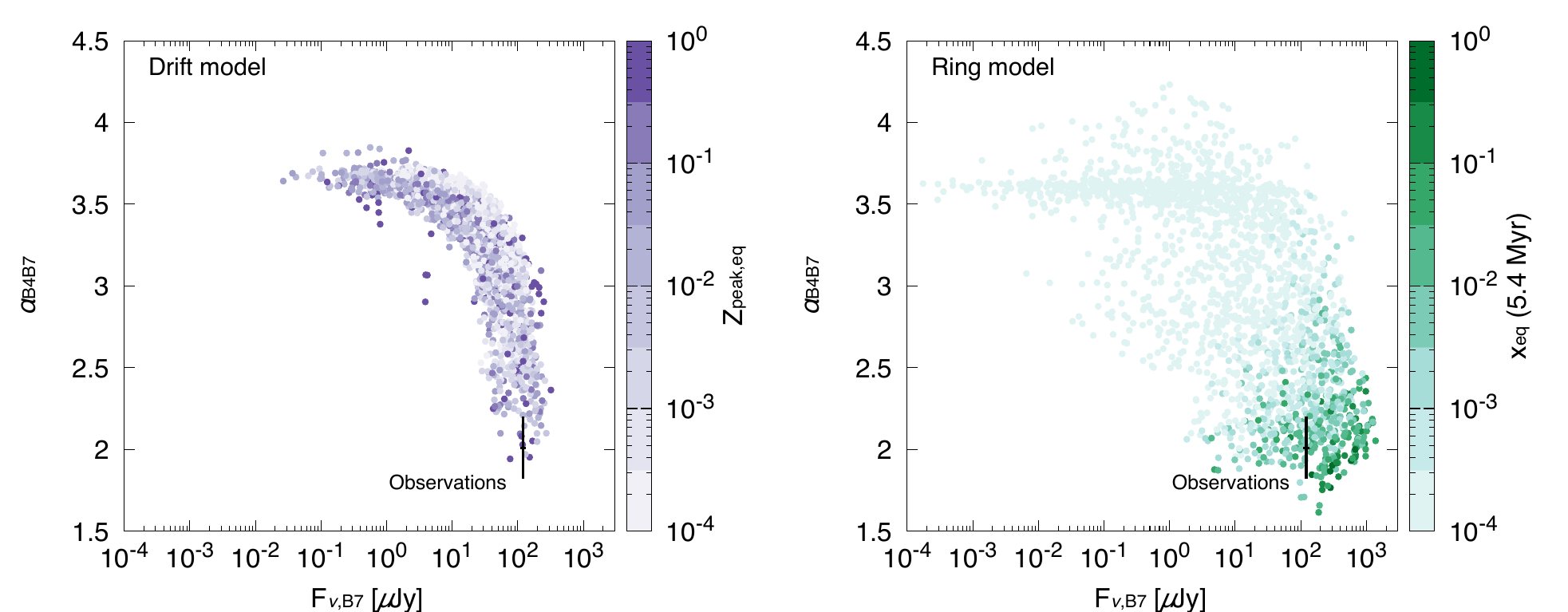}
\caption{Same as Fig. \ref{fig:comparisons}, but with colors representing different quantities. The colors in the left and right panels represent the peak dust-to-gas surface density ratio in the Drift model, $Z_{\rm peak,eq}$, and the equivalent dust-to-gas inflow mass ratio in the Ring model, $x_{\rm eq}$, respectively. The latter is calculated by equating the total dust mass in the CPD with the dust mass that would be supplied over $5.4~{\rm Myr}$ for $x_{\rm eq}$.
\label{fig:equivalent}}
\end{figure*}

We compare the dust-to-gas ratios in the Drift and Ring models, namely $x$ and $Z_{\rm peak}$, by defining equivalent quantities for each model. The left panel of Fig. \ref{fig:equivalent} represents the peak value of the dust-to-gas surface density ratio of the CPD of each calculation of the Drift model, $Z_{\rm peak,eq}$, which is equivalent to $Z_{\rm peak}$ of the Ring model. There is no clear relationship between the distribution of $Z_{\rm peak,eq}$ and that of $x$ (see the left panel of Fig. \ref{fig:comparisons}). The right panel of Fig. \ref{fig:equivalent} represents the dust-to-gas mass ratio in the gas inflow of the Ring model, $x_{\rm eq}$, which is equivalent to $x$ of the Drift model. Here, we derive the value from the assumption that the total dust mass in the CPD including the dust ring is equal to the dust mass that would be supplied to the CPD over $5.4~{\rm Myr}$ for $x_{\rm eq}$ (see Section \ref{sec:ring-disk}). The panel shows that the observed $\alpha_{\rm B4B7}$ and $F_{\rm \nu,B7}$ can be reproduced even when $x_{\rm eq}\lesssim0.1$, while the Drift model requires $x>0.1$ for the reproduction (see the left panel of Fig. \ref{fig:comparisons}).

\bibliography{CPD-Rings}{}
\bibliographystyle{aasjournalv7}



\end{document}